\documentclass[usenatbib]{mn2e}
\usepackage{graphicx}
\title[FUV Scattering by Dust in Orion]{FUV Scattering by Dust in Orion}
 \author[P. Shalima et al.]{P. Shalima$^{1}$ \thanks{E-mail:
shalima@iiap.res.in}, N. V. Sujatha$^{1}$ \thanks{E-mail:sujaskm@yahoo.com}, Jayant Murthy$^{1}$  \thanks{E-mail:jmurthy@yahoo.com},
Richard Conn Henry$^{2}$ \thanks{E-mail: henry@jhu.edu}
\and \hspace{7cm} and
\and \hspace{5.5cm} David J. Sahnow$^{2}$\thanks {E-mail:sahnow@pha.jhu.edu}\\
$^{1}$Indian Institute of Astrophysics, Bangalore 560034, India\\
$^{2}$Department of Physics and Astronomy, The Johns Hopkins University, Baltimore, MD 21218}
                                                                                
\begin{document}
\date{Accepted. Received; in original form}
                                                                                
\pagerange{\pageref{firstpage}--\pageref{lastpage}} \pubyear{}
                                                                                
\maketitle
                                                                                
\label{firstpage}
                                                                                
\begin{abstract}
We have modelled diffuse far-ultraviolet spectrum observed
by {\it FUSE} near M42 as scattering of starlight from the Trapezium stars
by dust in front of the nebula. The dust grains are known to be anomalous in 
Orion
with R$_{V} = 5.5$ and these are the first measurements of the FUV optical 
properties of the grains outside of ``normal'' Milky Way dust. We find an 
albedo varying from 0.3 $\pm$ 0.1 at 912 \AA\ to 0.5 $\pm$ 0.2 at 1020 \AA\, 
which is consistent with theoretical predictions.

\end{abstract}
                                                                                
                                                                                
\section{Introduction}

The Orion Nebula (M42) was first observed as one of the brightest
diffuse sources in the ultraviolet (UV) sky by \citet{Car77}. 
They identified this light to be due to the radiation from the
bright Orion stars scattered by dust in the Orion Molecular Cloud.
Further observations allowed \citet{Wen} to suggest a model for
the M42 region in which M42, itself, is a thin blister of ionized gas in
front of the Orion Molecular Cloud. The scattering arises in a neutral
sheet in front of the Nebula known as the Veil \citep{Odell92}
where there is a deficiency of small grains compared to the diffuse ISM \citep{Baade,Costero,Cardelli}.
 
The first observations of far ultraviolet 
(FUV: 905 -- 1187 \AA ) emission in this region were made by \citet{Mu05} who
used serendipitous 
pointings of the {\it Far Ultraviolet Spectroscopic Explorer} ({\it FUSE}) to 
find intensities as high as 3 $\times$ 10$^{5}$ ph cm$^{-2}$ s$^{-1}$ sr$^{-1}$ \AA$^{-1}$. 
In this work, we have modelled these observations in order to
determine the optical properties of the dust grains in the FUV. We find that 
the albedo of these grains varies from 0.3 $\pm$ 0.1 at 912~\AA\ to 0.5 $\pm$ 0.2 at 1020 \AA\ 
which is consistent with the theoretical predictions of \citet{Draine}. 

\section{Observations and Model}
\begin{table}
\caption[]{{\it FUSE} observations}
\label{fuse}
\begin{tabular}{ccccccccc}
\hline
Data set&l&b&Time&Intensity(1100\AA)\\
&(deg)&(deg)&(s)& ph cm$^{-2}$ s$^{-1}$ sr$^{-1}$ \AA$^{-1}$)\\
\hline
S4054601&208.8&-19.3&10565&2.93 $\pm$ 0.03 ($\times$10$^{5}$)\\
S4054602&208.8&-19.3&5696&2.95 $\pm$ 0.04 ($\times$10$^{5}$)\\
\hline
\end{tabular}
\end{table}

\begin{figure}
\centering
\includegraphics[height=9cm,width=9cm]{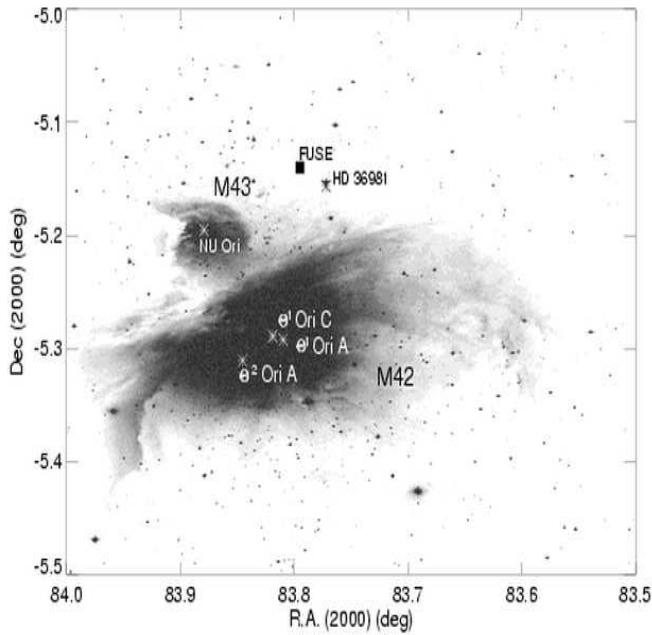}
\caption{DSS map of the region with the
 brightest stars (asterisks) and the {\it FUSE} location (filled square) overplotted.}
\label{lb}
\end{figure}

As part of the {\it FUSE} S405/505 program, regions of nominally blank sky were 
observed to allow the instrument to thermalize
before realignment of the spectrograph mirrors. \citet*{Mu04}
found diffuse radiation in many of these locations with the brightest being
the two in the vicinity of M42 \citep{Mu05}. The locations and brightnesses of these two
observations are listed in Table \ref{fuse} and the Digital Sky Survey (DSS) map
of M42 and the surrounding region is shown in Fig. \ref{lb} with the location of the
{\it FUSE} observation superimposed (square).

\begin{figure}
\centering
\includegraphics[height=9cm,width=9cm]{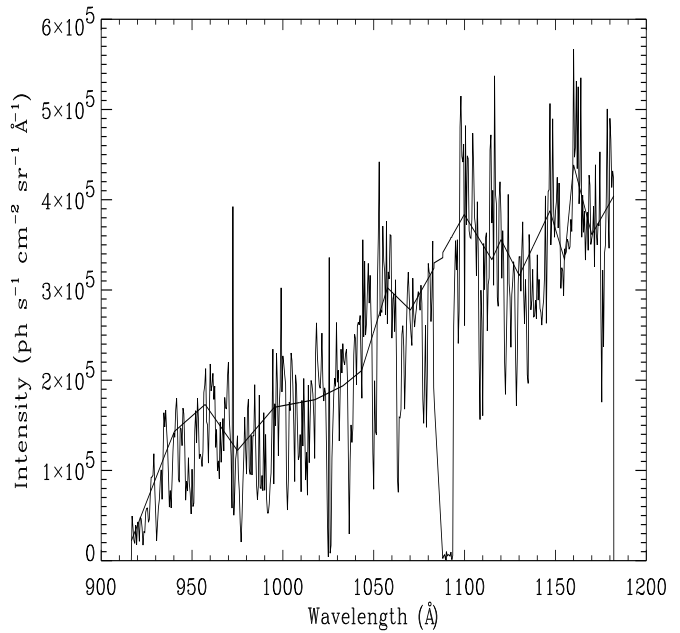}
\caption{Observed diffuse spectrum from \citet{Mu05}. The assumed dust
 scattering continuum is overplotted as a dark line.}
\label{spec_30}
\end{figure}

The spectrum of the diffuse light is shown in Fig. \ref{spec_30}. Although only 
1$\farcm$5 from HD 36981, \citet{Mu05} have shown that the 
observed emission could not be scattered radiation from that star because the 
broad photospheric Ly$\beta$ absorption line in the stellar spectrum is not
reflected in the diffuse spectrum and suggested that the radiation was instead
due to scattering of the light from the Trapezium cluster of stars. Indeed,
65\% of the radiation at the scattering location 
is provided by $\theta$$^{1}$ Ori C alone and 99\% 
by the four Trapezium stars in Table \ref{stars}.

\begin{table*}
\caption[]{Properties of stars near target}
\label{stars}
\begin{tabular}{ccccccccc}
\hline
HD No.&Name&$\it{l}$ &$\it{b}$ & angular distance&Sp. Type$^{{a}}$ &d$^{{b}}$ &Flux at location (1100 \AA) \\
 & &(deg) &(deg) &($\arcmin$)&&(pc) & ($\times$ 10$^{5}$ ph cm$^{-2}$ s$^{-1}$ \AA$^{-1}$)\\
\hline
37022&$\theta^{1}$ Ori C&209.01&-19.38&12.7&O6pe&450&1500\\
37041&$\theta^{2}$ Ori A&209.05&-19.37&14.8&O9.5V&450&370\\
37020&$\theta^{1}$ Ori A&209.01&-19.39&12.9&B0.5V&450&150\\
37023&$\theta^{1}$ Ori D&209.01&-19.38&12.7&B0.5Vp&450&250\\
\hline
\end{tabular}

{$^{a}$ obtained from the SIMBAD astronomical database}\\
{$^{b}$ Hipparcos catalogue}
\end{table*}
                                                                                
The amount of light scattered by the dust depends on the scattering properties 
of the interstellar dust grains (albedo, cross-section and scattering phase function), 
their number density and distribution, and the relative geometry of the stars 
and dust. We have used 
the Henyey-Greenstein function \citep{H.G.} for the scattering phase function. 
Although this is a purely empirical function which may not represent the true 
scattering, particularly for strongly forward scattering grains \citep{Draine},
it is the most prevalent in the literature.  

\vspace{3cm}
                                                                                
\begin{figure}
\centering
\includegraphics[height=8cm,width=8cm]{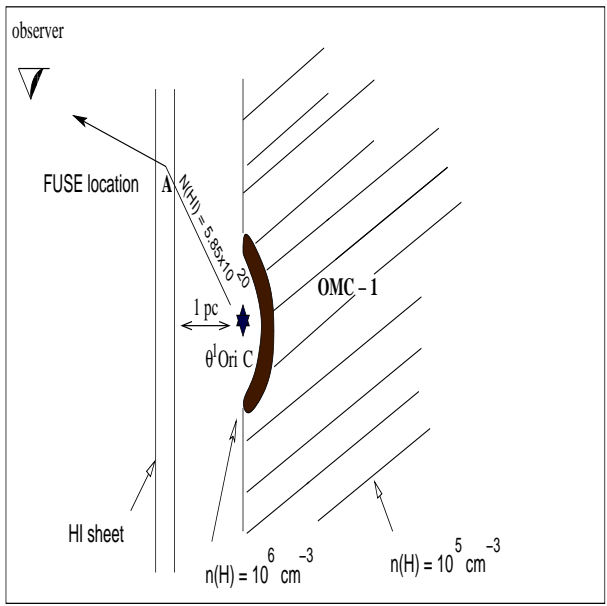}
\caption{Schematic representation of the distribution of dust at the
location showing the path (arrow) taken by the observed photons from
the Trapezium stars towards the observer. The figure is not to scale.}
\label{scheme}
\end{figure}

We have used the overall morphology of M42 and the surrounding medium described
in \citet{Odell01} to derive the scattering geometry (Fig. 
\ref{scheme}) for our location. Light from the Trapezium stars passes behind the foreground H {\small I} 
sheet (Orion's Veil) and is scattered by dust in the edges of the H {\small I} sheet at 
point A, about 3 $\pm$ 1 pc from $\theta$$^{1}$ Ori C. Assuming that the Veil
 is always parallel to the molecular cloud, this distance corresponds to the 
distance of the Veil \citep{Abel} at point A.

The strong interstellar absorption line at Ly$\beta$ indicates a total column 
density of N(H {\small I}) $=$ $(6.3 \pm 0.1) \times 10^{20}$ cm$^{-2}$ 
(using the ``H2OOLS'' packgage of \citet{McCandliss}) but 21~cm observations 
 \citep{Con98} show a much smaller column density of 4.5 $\times$ $10^{19}$ 
cm$^{-2}$ ($\tau_{912-1020~\mbox{\AA}}$ $\sim$ 0.029$-$0.033) at Point A \footnotemark[1]. 
\footnotetext[1]{The 21 cm intensities have been converted to H{\small I} column densities using the ratio of \citet{Werf} for material in the envelope of the molecular cloud.} Most of the absorption therefore arises in the 
medium between the Trapezium cluster of stars and the scattering location. 
Interestingly, the column density along the direct line of sight to the 
Trapezium stars (N(H) $=$ 3.9 $\times$ 10$^{21}$ cm$^{-2}$; \citet{shuping97}) is 
much higher than that seen in our scattered light observation indicating that 
we are actually observing light from $\theta$$^{1}$ Ori C reflected around the 
foreground clouds.

In order to convert the H {\small I} column densities into effective dust 
densities, we have used the dust cross-sections per hydrogen atom tabulated by
\citet{Draine} for an R$_{V}$ ($=A_{V}/E(B\hbox{-}V)$)~of 5.5, characteristic of the 
interstellar dust in Orion \citep{Cardelli89,Fitz99}. We note that
Draine's cross-sections assume the standard gas to dust ratio of 
\citet{Bohlin78}; however, this is 2.06 times higher than the ratio 
observed in Orion \citep{shuping97}. We have therefore reduced the dust 
cross-sections per hydrogen atom by this factor.
                                                                                
In summary, we have assumed that the observed emission is due to scattering of 
the light from the Trapezium stars by a scattering layer of thickness 
$4.5 \times 10^{19}$ cm$^{-2}$ at a distance of approximately 3 pc from $\theta$$^{1}$ Ori C. The 
radiation from the Trapezium stars has been attenuated by a column density of 
$6.3 \times 10^{20}$ cm$^{-2}$ using an extinction curve corresponding to 
R$_{V}$ = 5.5. 

The observed spectrum (Fig. \ref{spec_30}) includes many lines, both 
absorption and emission, of molecular hydrogen \citep{Fr05} as well as the H {\small I} 
Lyman series of absorption lines. In order to deduce the level of the dust 
scattered emission, we masked out these features and applied a 50 point median 
filter to the data (dark line in Fig. \ref{spec_30}). We then calculated the 
dust scattered radiation as a function of the optical constants $a$ and $g$ 
 using  single scattering (since $\tau$ $<$ 1) and compared the intensities with the observations to constrain the values of the parameters. 
Beyond 1020 \AA\ the observed radiation is contaminated by fluorescent emission of H$_{2}$ because of which we were unable to derive the optical constants. 

\section{Results and discussion}

\begin{figure}
\centering
\includegraphics[height=9cm,width=9cm]{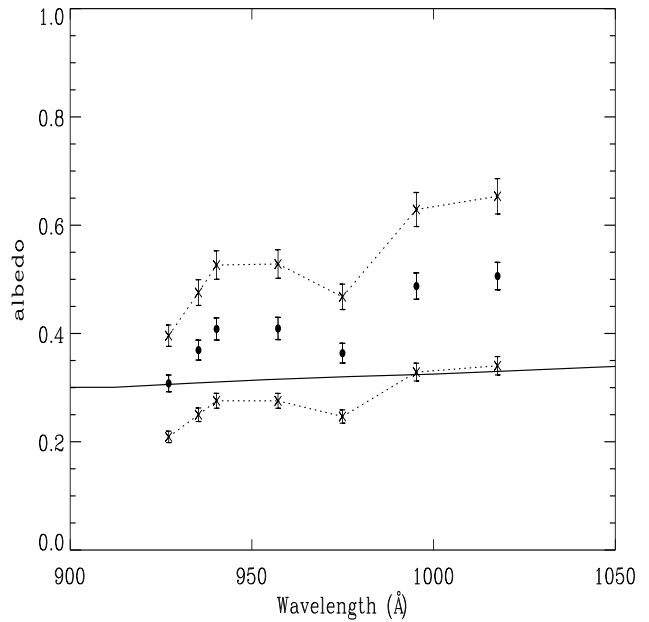}
\caption{Allowed values of $a$ (circles) corresponding to $g=0.55$, as a function of wavelength for dust cross-sections reduced by a factor of 2 compared to the R$_{V}$ $=$ 5.5 model. The minimum and maximum values corresponding to $g=0.4$ and $g=0.8$ respectively are marked as ``x''. The solid line represents the theoretical values of \citet{Draine}.}
\label{wavea}
\end{figure}

We have plotted our derived albedos as circles in Fig. \ref{wavea} corresponding to $g$ = 0.55. Because we
only have observations at a single location, where the scattering
angle is about $48\degr$, it was not possible for us to constrain $g$ and
 we assumed the value, $g=$0.55 $\pm$ 0.25 derived by \citet{Suj05} for 
locations in Ophiuchus. 
  The error bars on the albedos for each value of $g$, include 
 the observational errors as well as the uncertainties in the exact location of the scattering 
cloud. Within the assumed range of $g$, the albedo varies from 0.3 $\pm$ 0.1 at 
912 \AA\ to 0.5 $\pm$ 0.2 at 1020 \AA. The maximum allowed albedo corresponds to g=0.8 while the minimum corresponds to g=0.4. 

As mentioned above, the dust grains are anomalous in Orion with a depleted 
population of 
small grains with respect to the general interstellar medium perhaps due to 
the destruction and selective acceleration of the small grains by the stellar 
radiation combined with their coagulation into large grains \citep{Cardelli}. 
 However, the measured albedo is really a weighted average over all the 
different dust sizes and hence is only mildly dependent on R$_{V}$ in the FUV. 
In fact, we do find that our derived albedo is comparable to other determinations in the same wavelength region (Table \ref{albedo}). 

\citet{Cal} has derived
 the albedo and $g$ for IC 435 in Orion using data from the International 
Ultraviolet Explorer ({\it IUE}). They obtained a high albedo of 0.8 
corresponding to $g=0.75$ at 1200 \AA\ for dust with an R$_{V}$ of 5.3. However, as suggested by \citet{Burgh02}, if this is due to the low optical depths used by them, using the correct optical depths would result in low albedos similar
 to those obtained for other regions despite the difference in R$_{V}$.  

\begin{table}
\caption[]{Previous determinations}
\label{albedo}
\begin{tabular}{cccc}
\hline
Location&Wavelength&Albedo & $g$\\
&(\AA)&\\
\hline
NGC 7023&1000&0.42 $\pm$ 0.04&0.75\footnotemark[1]\\
NGC 2023&1100&0.35 $\pm$ 0.05&0.85\footnotemark[2]\\
Ophiuchus&1100&0.40$\pm$ 0.10&0.55 $\pm$ 0.25\footnotemark[3]\\
\hline
\end{tabular}

\thanks{\footnotemark[1] \citet{Witt93}}

\thanks{\footnotemark[2] \citet{Burgh02}}

\thanks{\footnotemark[3] \citet{Suj05}}

\end{table}

\section{Conclusions}
                                                                         
We have modelled the intense diffuse light observed near M42 by \citet{Mu05} as
starlight from the Trapezium stars scattered by 
interstellar dust in Orion's Veil, a sheet of neutral hydrogen in front of the 
Orion Nebula. Most of the absorption seen in the spectrum is due to material 
between the Trapezium and the scattering location and is much less than the 
absorption along the direct line of sight to the Trapezium. 

If we fix $g$ at 0.55, the albedo of the interstellar 
grains increases from a value of 0.3 $\pm$ 0.02 at 912 \AA\ to 0.5 $\pm$ 0.03 at 
1020~\AA\, close to previously observed values but with a different R$_{V}$. 
On the other hand, if we assume a $g$ of 0.85, as observed by \citet{Burgh02}, 
the albedo increases from 0.40 $\pm$ 0.02 to 0.68 $\pm$ 0.03. We have restricted our analysis up to 1020~\AA\ since molecular hydrogen
 fluorescence contaminates the spectrum longward of 1020~\AA. These 
observations are the first of dust grains with an R$_{V}$ significantly 
different from the Galactic norm in the FUV; however we do not find a 
significant difference in the optical properties.

\section*{acknowledgments}
We thank an anonymous referee for useful comments and suggestions which has
helped in improving the quality and presentation of the paper. 
We thank Kevin France for a careful reading of the manuscript and for useful suggestions. We also thank Dr. Sunetra Giridhar and Dr. P. Manoj 
for helpful discussions. This research has made use of the SIMBAD database,
operated at CDS, Strasbourg, France.
\bibliography{orion}
\label{lastpage}
\end{document}